\documentclass[%
 reprint,
superscriptaddress,
onecolumn,
amsmath,amssymb,
aps,
]{revtex4-2}

\usepackage{graphicx}% Include Figure files
\usepackage{color}
\usepackage{dcolumn}% Align table columns on decimal point
\usepackage{bm}% bold math
\usepackage{tikz}
\usepackage{siunitx}
\usepackage{float}
\usepackage[colorlinks = false]{hyperref}% add hypertext capabilities
\begin{document}

\title{Viscosity's impact on nutrient uptake along the gut}% Force line breaks with \\

\author{Fabian Karl Henn}
%\author{Agnese Codutti}%
\author{Karen Alim}
 \email{Contact author: k.alim@tum.de}
\affiliation{%
 \hyperlink{https://ror.org/02kkvpp62}{Technical University of Munich}, TUM School of Natural Sciences, Department of Bioscience,\\
Center for Protein Assemblies (CPA), 85748 Garching b.\ München, Munich, Germany
}%

\date{\today}% It is always \today, today,

\begin{abstract}
    Through switching contraction patterns driving digestive flows, the small intestine balances nutrient uptake and waste removal. Complex segmentation contraction patterns are associated with higher nutrient uptake, while peristaltic contractions primarily serve to flush out unused remnants. However, the impact of the Non-Newtonian behavior of digestive fluids on the efficacy of these contraction patterns remains unclear. Here, we present finite-element simulations that model nutrient transport, diffusion, and uptake within both segmentation and peristaltic contractions along the small intestine for Newtonian and Non-Newtonian fluids. Our simulations reveal that diffusion plays a key role in uptake, with nutrient absorption directly linked to fluid viscosity, which governs molecular diffusivity. Further, we present an analytical prediction for uptake in peristalsis as a function of molecular diffusivity, which aligns closely with our simulation results. Notably, we find that shear-thinning properties of Non-Newtonian fluids enhance nutrient uptake, particularly in segmentation contractions compared to peristalsis. Our results demonstrate the fluid dynamical principles underlying intestinal digestion, showing how shear-thinning Non-Newtonian fluids promote efficient nutrient uptake without compromising the clearance of waste. This physical insight advances our understanding of digestive processes and provides a foundation for exploring the prevention of intestinal diseases such as bacterial overgrowth. 
\end{abstract}

\maketitle

%\tableofcontents

\section{Introduction}	
		The phrase ``to have butterflies in your stomach" is a widespread synonym for a nervous feeling before a presentation, an important sports match or a lovely date. Of course it is not meant to be taken in a literal sense, however, stress or anxiety can indeed lead to an alteration in gut micro-biota \cite{AnxietyGut} and, thus, digestive problems. Surprisingly, the other way around is also possible, as it has been recently shown that enteroendocrine cells found in the thin epithelial sheet of the gut are capable of eliciting pain as well as stress-associated hormones and neurotransmitters \cite{GutAnxiety}. In consequence, a malfunctioning digestive system can not only lead to fatigue and debility but it can also be the cause of stress, anxiety and even depression, showing that research in this area is as important as ever. The small intestine, being the region where the majority of macronutrient uptake occurs, is of particular importance due to its critical role in nutrient absorption \cite{Macronutrientuptake}.\\
		There are two fundamental motor patterns exhibited by the small intestine, namely peristalsis and segmentation. Peristalsis facilitates the propulsion of contents through various biophysical systems, including the esophagus, ureter, male reproductive tract, small blood vessels, and the gastrointestinal tract. Mathematical models of peristaltic transport date back to fundamental theories based on the lubrication approximation and restricted to periodic, sinusoidal wave trains and infinitely long tubes \cite{Shapiro}. This framework was then extended for arbitrary wave shapes and finite length tubes \cite{Brasseur}. Further studies focused on the effect of the Reynolds number \cite{Takabatake1982, AYUKAWA1982, Takabatake1988}. In biological contexts, the transport of Non-Newtonian fluids such as blood, spermatozoa, or chyme in the digestive system has been extensively investigated \cite{Srivastava1984, Srivastava1985, SubbaReddy2007, velocityprofile, Hayat2004}. Peristalsis remains a prominent research topic with recent investigations focusing on wall shapes that mimic propagative membrane contractions \cite{Abokalesm}, the efficiency of peristaltic pumps with passive membrane components relevant to industrial applications \cite{Jensen}, and valveless pumping induced by phase shifts \cite{Amselem2023}. And yet the fluid physics describing how peristalitic contractions affect where particles diffusing in the flow are taken up along the intestine is unclear.\\ 
		Experimental recordings of the intestine have shown that peristalsis is active during inter-digestive periods and causes propulsion of remnant contents such as microbiota towards the anal direction while segmentation is activated upon food intake and promotes mixing and uptake of nutrients \cite{Originofsegmentation}. Furthermore, the interplay of available nutrients and bacteria may control which pattern is active trough a feedback loop suggesting that the depletion of nutrients triggers peristaltic cleaning to quickly reduce bacteria and prevent bacterial overgrowth whilst in return nutrient appearance triggers slower flows \cite{AgneseMainPaper}. Thus, to understand the digestive function of the small intestine both segmentation and peristalsis need to be compared side-by-side. Previous studies that focused on the effect of different contraction patterns \cite{Motilitycomparison} only considered local, non-propagating contractions and the resulting fluid mechanical consequences. Thus, a direct comparison of repeated peristaltic and segmental contractions on how they affect the motion and transport of single particles is yet needed.\\
		During the digestive process the rheological properties of the consumed food are vastly different depending on the considered stage of digestion, see Figure \ref{fig:tube geometry}. As food enters the proximal part of the intestine it is broken down into small food particles and intermixed with enzymes, bacteria and viscous digestive juices into a consistency referred to as chyme (Fig.~\ref{fig:tube geometry}, step 3). A large part of gastrointestinal research is focused on finding an adequate model to characterize the behavior of chyme during digestion \cite{Chymeimportance}. The main result that is agreed on from experimental studies is that chyme behaves like a Non-Newtonian fluid that exhibits shear thinning \cite{Shearthinningchickendigesta, Shearthinningpigdigesta} denoting that the apparent viscosity of chyme decreases at increased applied shear force. The reason for that is mostly considered to be the alignment of the long axes of the particles in the direction of the flow due to the shear forces. This results in less friction between adjacent particles as they are less likely to touch and, thus, a lower viscosity if the shear forces increase. The majority of experimental and theoretical investigations focus on different Non-Newtonian fluid models to approximate the chyme behavior. Models are based on the power-law fluid \cite{Studywithpowerlaw1, Studywithpowerlaw2} and extensions of it like the Herschel-Bulkley fluid \cite{Shearthinningchickendigesta} and the Carreau model \cite{Carreau}. Developing an adequate model to represent the rheological properties of chyme is crucial for advancing gastrointestinal research, however, due to the vast amount of parameters influencing chyme, no single model has been able to capture its full dynamic properties yet. Therefore, the power-law fluid with only two parameters  is the most widely used model.\\
		When studying the motion of macronutrients contained in chyme the effect of diffusion becomes of significance due to their minuscule radii. Consequentially many studies modeled the digestion and concentration changes of e.g.~glucose in the intestine with continuum convective-diffusion equations \cite{Convdiffusion1,Convdiffusion2,Convdiffusion3}. While continuum equations describe the development of how a macrosystem behaves, an investigation of single particle trajectories could help in grasping the fundamental mechanism driving individual nutrient particle uptake at the intestinal walls. Single particles subjected to peristaltic contractions have been studied before \cite{Jimenezlozano}. However, the particle motion was only governed by flow transport, but diffusion and uptake at the channel walls was lacking. Finally, the effect of a Non-Newtonian shear-thinning fluid, compared to a regular Newtonian fluid, on particles within intestine contraction driven flow has not been investigated.\\
		Therefore, in this work we numerically study particle motion within a Newtonian and Non-Newtonian fluid subject to intestinal contraction patterns to address how Non-Newtonian shear-thinning influences particle motion and uptake. In agreement with recent experimental \cite{PowerlawExperimental,Shearthinningchickendigesta, Shearthinningpigdigesta} and theoretical \cite{Studywithpowerlaw1,Studywithpowerlaw2} studies the Non-Newtonian fluid, referred to as chyme, will be modeled as a power-law fluid here, due to the simplicity of the equation allowing for analytic calculations. We distinguish the key differences between the two main intestinal contraction patterns, peristalsis and segmentation, with regard to particle transport and uptake location first before we address how fluid viscosity and shear-thinning affect uptake efficiency. We find that uptake is governed by molecular diffusivity through fluid viscosity and present an analytical prediction for uptake in peristalsis matching our numerical observations. Further, we reveal that a shear-thinning fluid promotes nutrient uptake in segmentation whilst at the same time retaining the important function of waste removal from peristalsis. 

\section{Methods}
	\subsection{Modelling the intestine}
    \subsubsection{The stages of digestion}\label{digestion outline}
        Upon intake at the mouth, food gets pushed by peristaltic contractions through the esophagus into the stomach as shown in step 1 in Figure \ref{fig:tube geometry} following Ref.~\cite{Sensoy2021}. In the stomach food is mixed with digestive enzymes and goes through a mechanical and chemical disintegration to form a fluidized mixture referred to as chyme in step 2, before being released by the pyloric sphincter into the small intestine in step 3. In the first segment of the small intestine, known as the duodenum, chyme is mixed with secretions from the pancreas, the gallbladder and glands in the intestinal wall such as enzymes, proteases, pancreatic lipase and amylase, step 4. In the second segment of the small intestine, the jejunum, the majority of nutrient uptake occurs, because here the villi, responsible for surface enlargement and thus uptake, are the largest, step 5. Remaining, useful substances are absorbed in the ileum which is the last segment of the small intestine, before the leftover chyme is transported into the large intestine, step 6. Here water is absorbed into the body and chyme turns back to a solid form known as feces. In the final step of digestion the formed feces is held in the rectum where it awaits defecation. 
	\subsubsection{Intestine contraction patterns}
\begin{figure}[h!]
     \centering
        \includegraphics[width = 0.95\columnwidth]{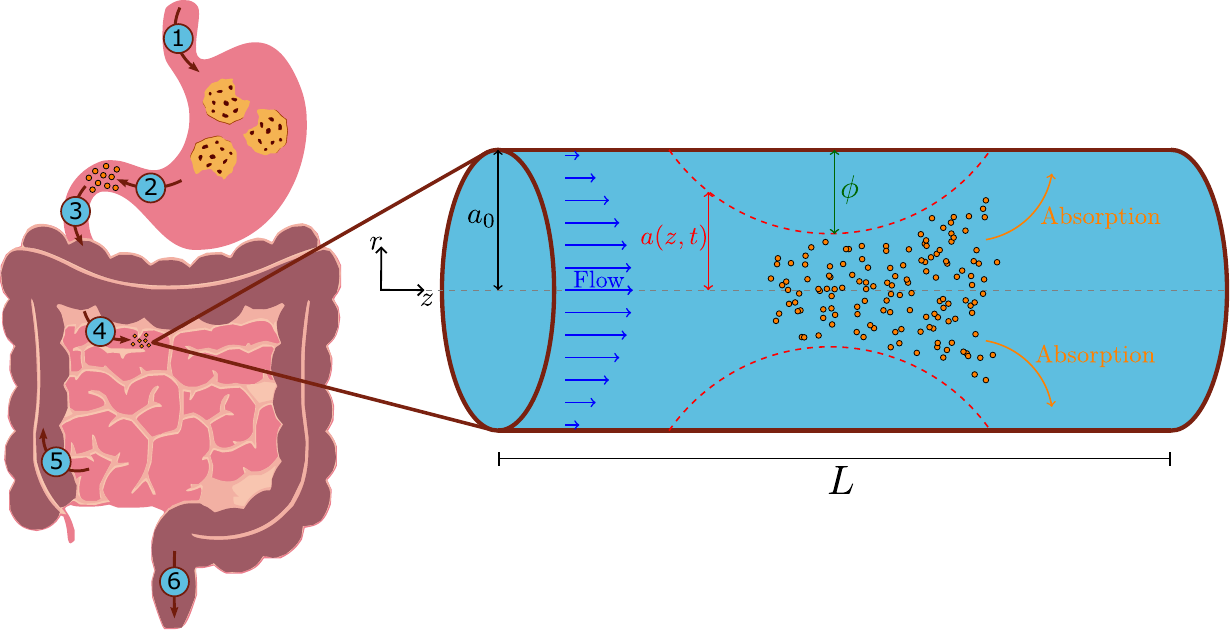}
		\caption{\textbf{Schematic diagram of the digestive system and the modeled tube geometry}\\
		The sketch on the left shows the transport way of food in the digestive system. 
        (1) Food gets pushed from the esophagus into the stomach. 
        (2) In the stomach food mixes with digestive enzymes and mechanical and chemical disintegration forms a fluidized mixture referred to as chyme which is released into the small intestine.
        (3) In the duodenum chyme is mixed with intestinal secretions. 
        (4) In the jejunum the majority of nutrient uptake happens,
        (5) Remaining, useful substances are absorbed in the ileum before chyme is transported to the large intestine.
        (6) Formed feces is held in the rectum until defecation.\\ 
        The cylinder on the right represents the modeled geometry. The rest radius is denoted as $a_0$ and the tube length as $L$. A possible contraction of the walls is indicated with the red dashed lines with the point of maximal occlusion marked as $\phi$ and the current radius in dependence of space and time with $a(z,t)$. The incoming parabolic velocity profile is visualized with the blue arrows where the length corresponds to the velocity magnitude. Exemplary nutrients that are advected by the flow are shown as small orange spheres that will be taken up upon touching the surrounding boundaries.}
		\label{fig:tube geometry}
	\end{figure}

    	The small intestine is modeled as a straight, smooth tube with constant radius that obeys azimuthal symmetry, see cartoon in Figure \ref{fig:tube geometry}. We define contraction patterns in cylindrical coordinates with the origin of the coordinate system  on the symmetry axis of the tube and on its base area. The cylinder length is denoted as $L = \SI{0.33}{\metre}$ \cite{MouseLength} and its rest radius as $a_0 = \SI{0.77}{\milli\metre}$ \cite{MouseRadius} to model a murine intestine. The study of animal intestines instead of a human one is widespread \cite{Intestinemodeloverview} due to a higher experimental coverage of the rheological properties of the contents as well as the intestinal contraction patterns. The spatio-temporal varying radius is labeled as $a(z,t)$. We further introduce the maximal occlusion $\phi = 1 - \max(\frac{a(z,t)}{a_0})$. Here $\phi = 1$ corresponds to a completely occluded tube where no flow is possible and $\phi = 0$ is equivalent to a straight tube. 

\begin{figure}[h!]
     \centering
        \includegraphics[width = 0.5\columnwidth]{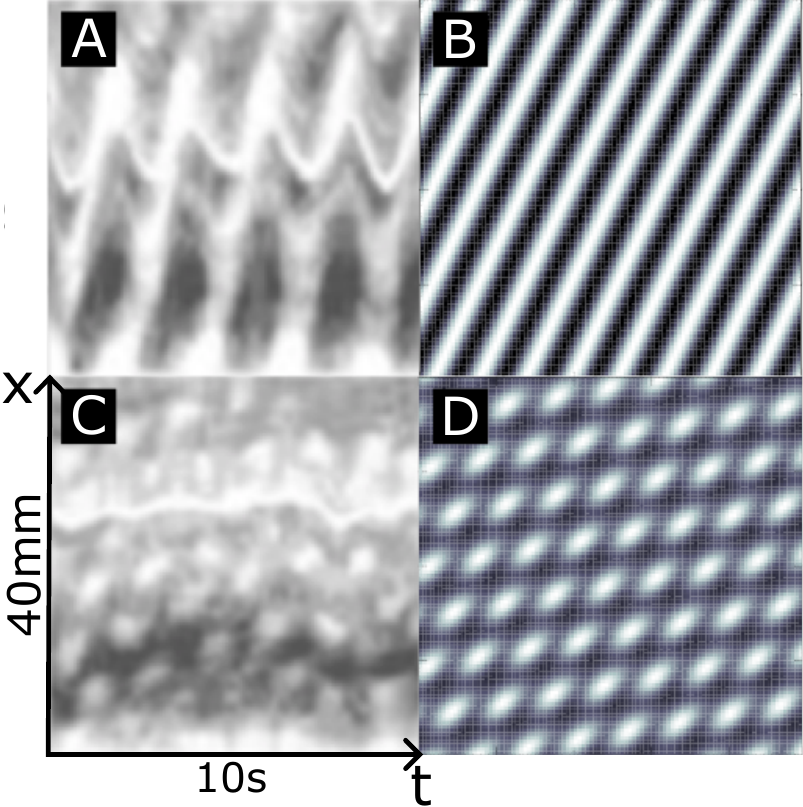}
		\caption{\textbf{Comparison of in vitro intestinal contractions with simulated contraction dynamics}\\
	In vitro spatiotemporal map of the contraction amplitude observed for the small intestine of mice, during peristalsis (a) and segmentation (c), respectively. Used with permission from Huizinga et al.\ ``Motor patterns of the small intestine explained by phase-amplitude coupling of two pacemaker activities: the critical importance of propagation velocity" \cite{Electroseg}. Simulated contraction amplitudes $\frac{a(t,z)}{a_0}$ with 10\% occlusion for peristalsis (b) and segmentation (d) capturing the experimentally observed contraction dynamics. Taken, with permission, from Codutti et al.\ ``Changing Flows Balance Nutrient Absorption and Bacterial Growth along the Gut" \cite{AgneseMainPaper}}
		\label{Electroplot}
	\end{figure}

	Our model of the contraction patterns was adapted from Codutti et al.~\cite{AgneseMainPaper} who obtained analytic expressions for both peristalsis and segmentation using the experimental results of Huizinga et al.~\cite{Originofsegmentation,Electroseg}. The latter studied the occurrence of segmentation and peristalsis in the jejunum of mice by electrophysiology. %In the following, the equations derived by Codutti et al.\ are introduced.
    For the peristaltic pattern the wall contractions follow a simple, continuously propagating wave train of the form  
	\begin{equation}
		\label{peristalsis}
		a(z,t) = a_0\left(1+\phi\sin\left(\Omega_\text{H} t - K_\text{H} z\right)\right),
	\end{equation}
	with oscillation frequency $\Omega_\text{H} = 2\pi\cdot\SI{0.83}{Hz}$, wave vector $K_\text{H} = 2\pi\cdot\SI{98}{m^{-1}}$ and the occlusion defined as the ratio of rest radius and wave amplitude $\phi = \frac{A_\text{P}}{a_0}$ where $A_\text{P}$ is adjusted to the desired occlusion. The corresponding wavelength and wave period are $\lambda_\text{P} = \SI{0.0102}{\meter}$ and $T_\text{P} = \SI{1.205}{\second}$.\\
	Segmentation is modeled by the superposition of a low and a high frequency wave combined with a carrier and an envelope wave as 
	\begin{equation}
		\label{segmentation}
		a(z,t) = a_0\left(1+\phi \frac{\phantom{\|}\xi\phantom{\|}}{\|\xi\|}_\text{max}\right),
	\end{equation}
    with
	\begin{equation}
        \begin{split}
		\label{segmentation signal}
		\xi &=\Gamma_\text{H}\sin\left(\Omega_\text{H} t - K_\text{H} z\right) + \Gamma_\text{L} \sin\left(\Omega_\text{L} t - K_\text{L} z\right)\\
		&+ \frac{\Gamma_\text{P}}{2} (\sin\left( (K_\text{L} - K_\text{H}) z + (\Omega_\text{H} - \Omega_\text{L}) t + \theta \right) - \sin\left((K_\text{H} + K_\text{L}) z + \theta - (\Omega_\text{H} +  \Omega_\text{L}) t\right)).	
	   \end{split}
    \end{equation}
	The parameters for the high-frequency wave $\Omega_\text{H}$ and $K_\text{H}$ are the same as for the peristaltic case, for the low-frequency wave $\Omega_\text{L} = 2\pi\cdot\SI{0.13}{\hertz}$ and $K_\text{L} = 2\pi\cdot\SI{190.5}{m^{-1}}$, the wave amplitudes are $\Gamma_\text{H} = 0.78$, $\Gamma_\text{L} = 1$, $\Gamma_\text{P} = 0.48$, respectively, and the phase shift of the coupling term is $\theta = 2\pi \cdot \SI{0.3}{rad}$. The normalization by the maximum value of the oscillating term is necessary to maintain the definition of occlusion $\phi$ as the percentage of radius change and to ensure that the maximal contraction/expansion of the tube is $a_0(1-\phi)<a<a_0(1+\phi)$. All values for both peristalsis and segmentation are in accordance with the model of Codutti et al.~\cite{AgneseMainPaper} and were derived from the experimental results of Huizinga et al.\ \cite{Originofsegmentation,Electroseg}. See Fig.~\ref{Electroplot} for an comparison of in vitro spatio-temporal maps of the contraction amplitude observed for the small intestine of mice during peristalsis and segmentation and simulated contraction
	amplitudes based on Eqs.~\eqref{peristalsis}- \eqref{segmentation signal}. The exact wall shape along a two-dimensional longitudinal cross-section is depicted in Fig.~\ref{fig:perist vs seg}a and \ref{fig:perist vs seg}b for peristalsis and segmentation, respectively. Throughout this work we only consider an occlusion of $\phi = 0.95$ as the magnitude of all phenomena discussed in the following increases with increasing occlusion.
 
	\subsubsection{Rheological model of the chyme fluid}
	The constitutive equation of the chyme fluid, in contrast to water as our reference Newtonian fluid of constant viscosity, is described by a power-law often called the Ostwald–de Waele power-law model with the shear stress tensor $\boldsymbol{\tau} = -\eta \boldsymbol{\dot{\gamma}}$. Here $\boldsymbol{\dot{\gamma}} = \mathbf{\nabla u} + \left(\mathbf{\nabla u}\right)^\intercal$ corresponds to the rate of strain tensor with the velocity field given by $\mathbf{u}$. The viscosity function $\eta = m \dot{\gamma}^{n-1}$ depends on the ``magnitude of the rate of strain tensor'' defined by the double inner product $a:b = a_{ij}b_{ij}$ as $\dot{\gamma} = \sqrt{\frac{1}{2}(\boldsymbol{\dot{\epsilon}}:\boldsymbol{\dot{\epsilon}})}$ as well as the flow consistency index m and the flow behavior index n. The value for $n$ is taken from pig digesta of the proximal intestine as $n = 0.252$ \cite{Shearthinningpigdigesta} and $m$ is chosen such that the average viscosity is the same as water allowing for a meaningful comparison between the two fluid types. Thus, $m = \SI{0.014}{\pascal\second}^n$ will be used for the chyme model.\\
	The particles inside the chyme are considered separately as non-interacting small spheres characterized by the diffusion coefficient $D = 0.5\cdot10^{-11}\SI{}{\meter\squared\per\second}$ of fatty acids \cite{FattyAcidDiffusion} found in the small intestine \cite{FattyAcid}. The diffusion coefficient is given by $D = \frac{k_\text{B}T}{6\pi\mu r_0}$ where $k_\text{B}$ is the Boltzmann constant, $T$ is the absolute temperature in Kelvin and $\mu$ the dynamic viscosity of the solute. By rearranging and inserting the viscosity of water, a particle radius of $r_0 = 4.29\cdot10^{-8}\SI{}{\meter}$ is calculated which will be used as a constant radius for all simulated particles independent of the chosen viscosity. Owing to the minuscule radius of the particles, forces such as buoyancy and gravity, which are proportional to the particle volume, are neglected in the equation of motion. Thus, only the drag force due to flow and white noise forcing resulting in particle diffusion are included. However, with the described particle parameters the equation of motion exhibits severe numerical stiffness causing the solver to compute multiple days for just a few milliseconds of simulation. Therefore, the equation of motion was simplified by assuming that the particles instantaneously reach their terminal velocity in each time step, i.e.~inertia can be ignored. This is justified as the Lagrangian time scale  $\tau_P = \frac{\rho_{p}r^2}{72\mu} \sim \mathcal{O}\left(10^{-11}\right)\SI{}{s}$ with particle density $\rho_{p}$, radius $r$ and viscosity $\mu$ is significantly lower than the typical time scale of the system given by $\frac{\lambda}{c} \sim \mathcal{O}\left(1\right)\SI{}{s}$ with the wave speed $c$. Thus, the simplified equation of motion for a particle of mass $m_P$ subjected to drag from a velocity field $\mathbf{u}$ and white noise $\mathbf{F_{W}}$ is 
	\begin{equation}
        \label{Particle EOM}
		\dot{\mathbf{x}} = \frac{\tau_P}{m_P}\mathbf{F_{W}} + \mathbf{u}.
	\end{equation}
	\subsection{Flow Profile}\label{Flow velocity field}
	To describe the fluid flow ensuing from contraction patterns here the Navier-Stokes equations are solved. The derivation is based on  Misra and Maiti \cite{velocityprofile}.
	To apply the lubrication approach the variables are non-dimensionalized and terms of order $Re$ and $\delta$ are neglected.
	Here $Re$ is the Reynolds number for a power-law fluid, given by $Re = \frac{\rho c^{2-n} a_0^n\delta}{m}$ with the fluid density $\rho$, the wave speed $c$ and the lubrication parameter $\delta = \frac{a0}{\lambda}$. This results in the following equations
	\begin{equation}
		\label{Simplified Navier-Stokes}
		\frac{\partial p}{\partial r} = 0, \phantom{lol} \frac{\partial p}{\partial z} = \frac{m}{r} \frac{\partial }{\partial r}\left(r\left|\frac{\partial u}{\partial r}\right|^{n-1}\frac{\partial u}{\partial r} \right),
	\end{equation}
	combined with zero pressure in- and outflow boundary conditions $p(z = 0) = p(z = L) = 0$. At the tube center we impose the symmetry of the flow profile $\frac{\partial u}{\partial r}|_{r = 0} = 0$ and $v(r = 0) = 0$ and the no-slip condition at the tube walls $u(r = a) = 0$ and $v(r = a) = \frac{\partial a}{\partial t}$.
	Solving Eq.~\eqref{Simplified Navier-Stokes} for the axial velocity results in 
	\begin{equation}
		\label{axial velocity}
		u = \frac{n}{(2m)^{\frac{1}{n}}(n+1)} \frac{\partial p }{\partial z} \left|\frac{\partial p }{\partial z}\right|^{\frac{1-n}{n}} \left(r^{\frac{n+1}{n}} - a^{\frac{n+1}{n}}\right),
	\end{equation}
	with the continuity equation for incompressible fluids $\vec{\nabla}\cdot\vec{v} = 0$ the radial velocity is determined
	\begin{eqnarray}
		\label{radial velocity}
		v = \frac{1}{(2m)^{\frac{1}{n}}(n+1)}\left|\frac{\partial p }{\partial z}\right|^{\frac{1-n}{n}}
        \left(\frac{\partial^2 p }{\partial z^2}\left(\frac{1}{2}ra^{\frac{n+1}{n}} - \frac{n}{3n+1}r^{\frac{2n+1}{n}}\right)
        + \frac{n+1}{2}\frac{\partial p }{\partial z}ra^{\frac{1}{n}}\frac{\partial a}{\partial z}\right).
	\end{eqnarray}
	Using the no-slip condition $v(r = a) = \frac{\partial a}{\partial t}$ yields and equation for the pressure gradient 
	\begin{eqnarray}
		\label{pressure gradient}
		\frac{\partial p }{\partial z} =\frac{2^{n+1}m\left(\frac{3n+1}{n}\right)^n}{a^{3n+1}}\left(ca_0^2G(t) + \int_{0}^{z}dz' a\frac{\partial a}{\partial t}\right)\left|ca_0^2G(t) + \int_{0}^{z}dz' a\frac{\partial a}{\partial t}\right|^{n-1}.
	\end{eqnarray}
	Which simplifies for $n = 1$ to results found by Li and Brasseur \cite{Brasseur}. Here $G(t)$ is the integration function which at most depends on time. Integrating the pressure gradient gives an expression for the function $G(t)$ as a function of the applied pressure drop $\Delta p$ along the tube
	\begin{eqnarray}
		\label{Integration function}
		\Delta p = \int_{0}^{z}dz' \frac{\partial p}{\partial z'} = 2^{n+1}m\left(\frac{3n+1}{n}\right)^n
        \int_{0}^{z} dz' \frac{1}{a^{3n+1}}\left(ca_0^2G(t) + \int_{0}^{z'}dz'' a\frac{\partial a}{\partial t}\right)
        \left|ca_0^2G(t) + \int_{0}^{z'}dz'' a\frac{\partial a}{\partial t}\right|^{n-1},
	\end{eqnarray}
	which is then solved numerically. Finally, using the boundary condition $v(r = a) = \frac{\partial a}{\partial t}$ gives an expression for the derivative of the pressure gradient
	\begin{eqnarray}
		\label{2nd pressure gradient}
		\frac{\partial^2p}{\partial z^2} = \frac{3n+1}{a^{\frac{1+2n}{n}}}
        \left((2^{n+1}m)^{\frac{1}{n}}\left|\frac{\partial p}{\partial z}\right|^{\frac{n-1}{n}}\frac{\partial a}{\partial t}- \frac{\partial p}{\partial z}a^{\frac{n+1}{n}}\frac{\partial a}{\partial z}\right).
	\end{eqnarray}
For validation of our finite-element flow simulations with COMSOL in Fig.~\ref{fig:velocity_profile} we solve for the flow field in the lubrication approximation  by first determining $G(t)$ from Eq.~\eqref{Integration function}, then inserting $G(t)$ into Eq.~\eqref{pressure gradient} to calculate the pressure gradient and finally using the pressure gradient to determine the axial velocity $u$ and radial velocity $v$ with Eqs.~\eqref{axial velocity},\eqref{radial velocity}.
Although an analytical expression for the flow field is available, numerically calculating the flow field for the entire tube for around $\SI{200}{\second}$ is tedious and prone to numerical errors due to the highly oscillatory behavior of the integrand in Eq.~\eqref{Integration function}. Therefore, we turn to the powerful software COMSOL Multiphysics which allows us to obtain much quicker and more accurate solutions. Additionally, COMSOL includes many possibilities to incorporate random white noise as required by Eq.~\eqref{Particle EOM} for particle trajectories.
	\subsection{Numerical implementation}
\begin{figure}[h!]
	   	    \includegraphics[width = 0.8\columnwidth]{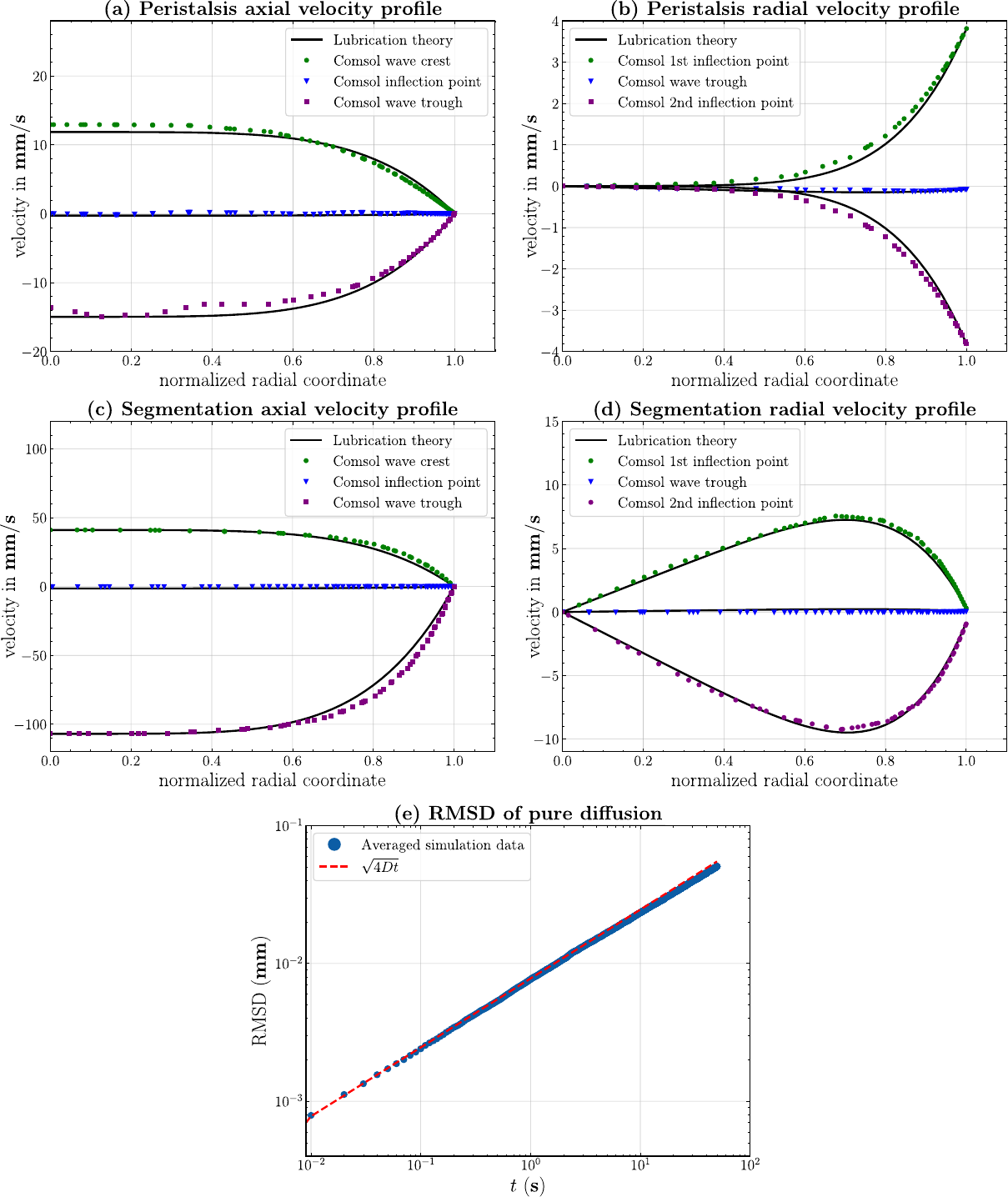}
            \caption{\textbf{Comparison of results from COMSOL with analytical calculations.}\\
    		Axial velocity profile $u(r,z,t)$ for peristalsis (a) and segmentation (c) as function of the normalized radial coordinate at three different longitudinal positions along the contraction pattern, the wave crest, the inflection point an the wave trough.
    		Radial velocity profile $v(r,z,t)$ for peristalsis (c) and segmentation (d).
            (e) Averaged root mean squared displacement of all particles over time}
    		\label{fig:velocity_profile}
		\end{figure}

		All simulations are performed using the software COMSOL Multiphysics which applies the finite element method to solve the full, incompressible Navier-Stokes equations for the fluid flow and particle motion according to Eq.~\eqref{Particle EOM}. Due to the azimuthal symmetry of the intestine the fluid flow and the particle movement was computed in a two dimensional axis-symmetric geometry. Here the no-slip condition was used at the wall boundaries and the pressure at the in- and outlet of the tube were set to zero. The tube deformations are implemented by the ``prescribed mesh displacement'' node to account for the pressure gradient caused by the intestinal contraction. For a flow behavior index $n<1$ the apparent viscosity diverges if the shear rate approaches zero, which is why a lower bound for the shear rate is specified to prevent numerical instabilities. In all simulations a value of $\dot{\gamma}_{min} = 10^{-10}\SI{}{\frac{1}{\second}}$ was chosen to achieve a high accuracy.\\
		In order to validate the velocity field computed by COMSOL the results are compared with the axial and radial velocity field defined by Eqs.~\eqref{axial velocity} and \eqref{radial velocity}. The corresponding axial and radial velocity profiles are shown in Fig.~\ref{fig:velocity_profile}. Three characteristic times of forward, backward and resting flow at the wave crest, wave trough and inflection point of the wall shape are compared with good agreement between lubrication theory and COMSOL solutions. Small deviations arise likely due to the neglected terms from the lubrication assumption. Moreover, the accurate implementation of the particles' diffusive motion is validated by calculating the average root mean squared displacement (RMSD), defined as $RMSD = \frac{1}{N}\sum_{i = 1}^{N}\sqrt{(x(t-t_0)-x(t_0))^2}$. A simulation without wall contractions was conducted, and the resulting RMSD, shown in Fig.~\ref{fig:velocity_profile}e confirms the expected proportionality to  $\sqrt{t}$.\\
		Particle uptake is modeled to take place whenever a particle comes in contact with the wall, i.e.~we assume a 100\% uptake probability. This is in contrast to absorption dynamics described by reaction diffusion equations, where the uptake absorption rate corresponds to a less than 100\% uptake probability \cite{Particledispersion}. However, regarding the scope of our study the uptake probability would not change the dynamics of the system but only decrease the total uptake numbers, diminishing visibility of mechanisms driving uptake.
		To initialize simulations particles are evenly distributed over one wavelength for peristalsis and a corresponding distance for segmentation. The total particle count is $3784$ for peristalsis and $3815$ segmentation. All variables, their value and a corresponding reference are shown as an overview in table \ref{tab:parameters}.
    	\begin{table}[h!]
			\centering
			\caption{Parameters and values used in the analysis.}
			\begin{tabular}{|l|l|l|l|}
				\hline
				\textbf{Variable} 	& \textbf{Symbol} 	& \textbf{Value}& \textbf{Reference} \\ 
				\hline
				Length of tube 		& $L$ 				& $\SI{0.33}{\meter}$			& Murine small intestine length \cite{MouseLength}\\ 
				Rest radius			& $a_0$ 			& $\SI{0.77}{\milli\meter}$ 	& Murine small intestine radius \cite{MouseRadius}\\
				Occlusion 			& $\phi$ 			& $0.95$				 		& Human occlusion is around $0.75$ \cite{HumanOcclusion}\\ 
				High-frequency wave frequency  	& $\Omega_H$ & $2\pi \cdot \SI{0.83}{\hertz}$ 		& \cite{AgneseMainPaper},\cite{Electroseg}\\ 
				High-frequency wave wavelength 	& $K_H$ 	 & $2\pi \cdot \SI{98}{\per\meter}$ 	& \cite{AgneseMainPaper},\cite{Electroseg}\\ 
				Low-frequency wave frequency 	& $\Omega_L$ & $2\pi \cdot \SI{0.13}{\hertz}$ 		& \cite{AgneseMainPaper},\cite{Electroseg}\\ 
				Low-frequency wave wavelength 	& $K_L$ 	 & $2\pi \cdot \SI{190.5}{\per\meter}$ 	& \cite{AgneseMainPaper},\cite{Electroseg}\\ 
				High-frequency wave amplitude 	& $\Gamma_H$ & $0.78$ 					& \cite{AgneseMainPaper},\cite{Electroseg}\\ 
				Low-frequency wave amplitude 	& $\Gamma_L$ & 1 						& \cite{AgneseMainPaper},\cite{Electroseg}\\ 
				Modulating wave amplitude 		& $\Gamma_P$ & 0.48 					& \cite{AgneseMainPaper},\cite{Electroseg}\\ 
				Phase shift 					& $\theta$ 	 & $2\pi \cdot 0.3$ 		& \cite{AgneseMainPaper},\cite{Electroseg}\\
				Peristaltic wavelength			& $\lambda_P$& \SI{0.0102}{\meter} 		& Phase velocity $v_P = 0.85\cdot 10^{-2}\SI{}{\meter\per\second}$ \cite{Electroseg}\\
				Segmental wavelength			& $\lambda_S$& $\SI{0.357}{\meter}$		& Phase velocity $v_S = \SI{0.119}{\meter\per\second}$\cite{AgneseMainPaper},\cite{Electroseg}\\
				Peristaltic wave period			& $T_P$		& $\SI{1.2}{\second}$		& Phase velocity $v_P = 0.85\cdot 10^{-2}\SI{}{\meter\per\second}$ \cite{Electroseg}\\
				Segmental wave period			& $T_S$		& $\SI{30.0}{\second}$		& Phase velocity $v_S = \SI{0.119}{\meter\per\second}$\cite{AgneseMainPaper},\cite{Electroseg}\\
				Viscosity of water				& $\mu$ 	 & $\SI{0.001}{\pascal\second}$ & \cite{ViscosityWater}\\
				Flow behavior index of chyme model	& $n$ 	 & $0.252$ 		& Pig digesta in proximal intestine \cite{Shearthinningpigdigesta}\\ 
				Consistency index of chyme model & $m$  & $\SI{0.014}{\pascal\second}$ & Results in same average viscosity as water\\  
				Fixed pressure drop 			& $\Delta p$ & \SI{0}{\pascal} 			& Pure pumping due to contraction \cite{AgneseMainPaper}\\ 
				Diffusion coefficient 			& $D$ 		 & $0.5\cdot 10^{-11}\SI{}{\meter\squared\per\second}$& Diffusivity of fatty acids \cite{FattyAcidDiffusion}\\ 
				Particle radius					& $r_0$		 & $4.29\cdot10^{-8}\SI{}{\meter}$ & Calculated from Diffusion coefficient\\
				Minimal shear rate				& $\dot{\gamma}_{min}$ & $10^{-10}\SI{}{\per\second}$ & Numerical accuracy \\
				Total particle number peristalsis	& $N_P$ & $3784$ & Uniform distribution over one wavelength \\
				Total particle number segmentation	& $N_S$ & $3815$ & Uniform distribution over equivalent length\\
				\hline
			\end{tabular}
			\label{tab:parameters}
		\end{table}
\section{Results}
	To start we analyze the uptake characteristics of peristalsis and segmentation for a Newtonian fluid of water's viscosity. This allows us later on to assess the impact of the Non-Newtonian shear-thinning of chyme on the uptake dynamics. 
	\subsection{Peristalsis transports particles in bolus whereas segmentation causes spatially distributed uptake}
\begin{figure}[h!]
        \includegraphics[width = 0.8\columnwidth]{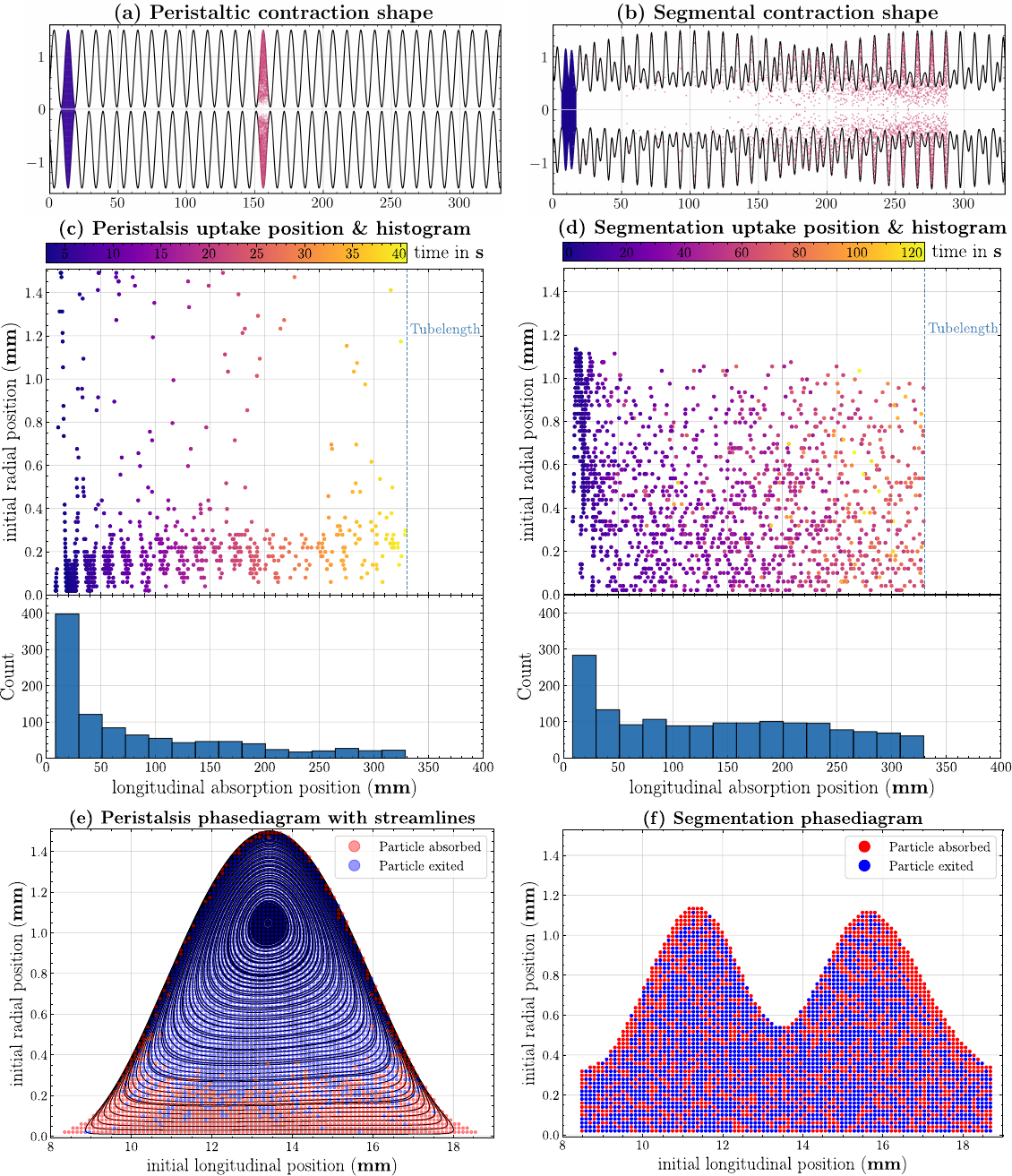}
		\caption{\textbf{Peristaltic uptake is governed by bolus formation leading to rapidly decaying uptake along the tube whilst in segmentation mixing is enhanced and uptake is more spatially distributed.}\\
		Two-dimensional tube cross-section visualizing simulated particles at time $t=\SI{0}{\s}$ and $t=\SI{20}{\s}$ for (a) peristaltic contraction and at $t=\SI{0}{\s}$ and $t=\SI{60}{\s}$ for (b) segmental contraction patterns. The longitudinal uptake position of particles in dependence of their initial radial position for peristalsis (c) and segmentation (d) reveals a direct correlation between for peristalsis, resulting in less and less particles being taken up along the tube as shown by the corresponding histogram below. On the contrary for segmentation no radial-longitudinal correlation exists and particles get after an initial absorption peak roughly homogeneously taken up along the tube as visualized by the corresponding histogram. Color-coding of which particles are taken up (red) or exit the tube (blue) mapped onto the initialization time point for peristalsis (e) and segmentation (f) reveals for peristalsis that mainly particles initially located within the outer streamlines get taken up while particles mix in segmentation. Black lines: Overlay of peristalsis streamlines .}
		\label{fig:perist vs seg}
	\end{figure}

		Subjected to peristaltic contractions, particles are trapped in a bolus that travels with wave speed through the tube, see Fig.~\ref{fig:perist vs seg}a for example at $t\approx \SI{20}{\second}$. This is the result of the mainly circling streamlines in the wave frame for a single wavelength, see Fig.~\ref{fig:perist vs seg}e. At the high occlusion considered here, the tube is almost closed after each wavelength, thus, stopping almost all particles from escaping their initial placement in the contracted wall segment. Mapping out where particles get taken up and when as a function of their initial position, see Fig.~\ref{fig:perist vs seg}c, we observe that with increasing time the longitudinal uptake position $l_{\rm{abs}}$ increases meaning that particles are taken up further down in the tube. In addition particles with a small initial radial position $r_0$, i.e.~close to the tube center, are generally taken up earlier than particles that started farther away from the tube center. This observation can be explained by the circling streamlines, see Fig.~\ref{fig:perist vs seg}e. Particles starting close to the center line circle on outer streamlines close to the tube wall over the course of a wave and, thus, are likely to diffuse to the tube wall and be taken up. Instead, particles entrapped in the bolus at inner streamlines require more time to diffuse to the outer streamlines circling close to the tube wall to eventually come in contact with the wall to be taken up. 
        Additionally, with each subsequent contraction wave fewer and fewer particles are taken up as the probability to diffuse a further distance decreases and the majority of the remaining particles accumulate near the center of the bolus which enhances their longitudinal transport and prevents them from being taken up. This can be seen in the decreasing uptake count over the tube length, see histogram in Fig.~\ref{fig:perist vs seg}c and the initial position of the particles color-coded with their uptake time in Fig.~\ref{fig:perist vs seg}e. Here, the higher the initial radial coordinate, the lower the amount of particles that get taken up.\\ 
		In segmentation particles are sloshed back and forth due to the superposition of waves with different wave speeds and amplitudes leading to particle mixing, see Fig.~\ref{fig:perist vs seg}b. Particles uptake location is much more spread out along the tube axis as compared to peristalsis, where particles are trapped in a bolus. In addition, particles spend comparably much longer time in the tube under segmentation, as shown in Fig.~\ref{fig:perist vs seg}d, as a consequence uptake also happens at much later times close to the tube inlet in contrast to the direct correlation of uptake time and distance along the tube in peristalsis. Thus, with segmentation overall uptake is promoted due to the longer residence time of the particles increasing their probability to be taken up and also the uptake is spread out along the tube for all times instead of the localized uptake in dependence of the bolus position for peristalsis. Additionally, the uptake count along the tube remains, apart from the initialization peak, approximately constant showing that as long as particles remain in the tube they will be taken up in a finite time as opposed to peristalsis where this probability decreases over time. Furthermore, uptake is independent on the initial radial position of the particles, see Fig.~\ref{fig:perist vs seg}f. Our findings, thus, underline that segmentation enhances the mixing of particles and promotes uptake of nutrients independent of their initial position in contrast to the selective uptake in peristalsis.
	
	\subsection{Diffusive streamline hopping causes uptake in peristalsis}
\begin{figure}[h!]
        \includegraphics[width = 0.8\columnwidth]{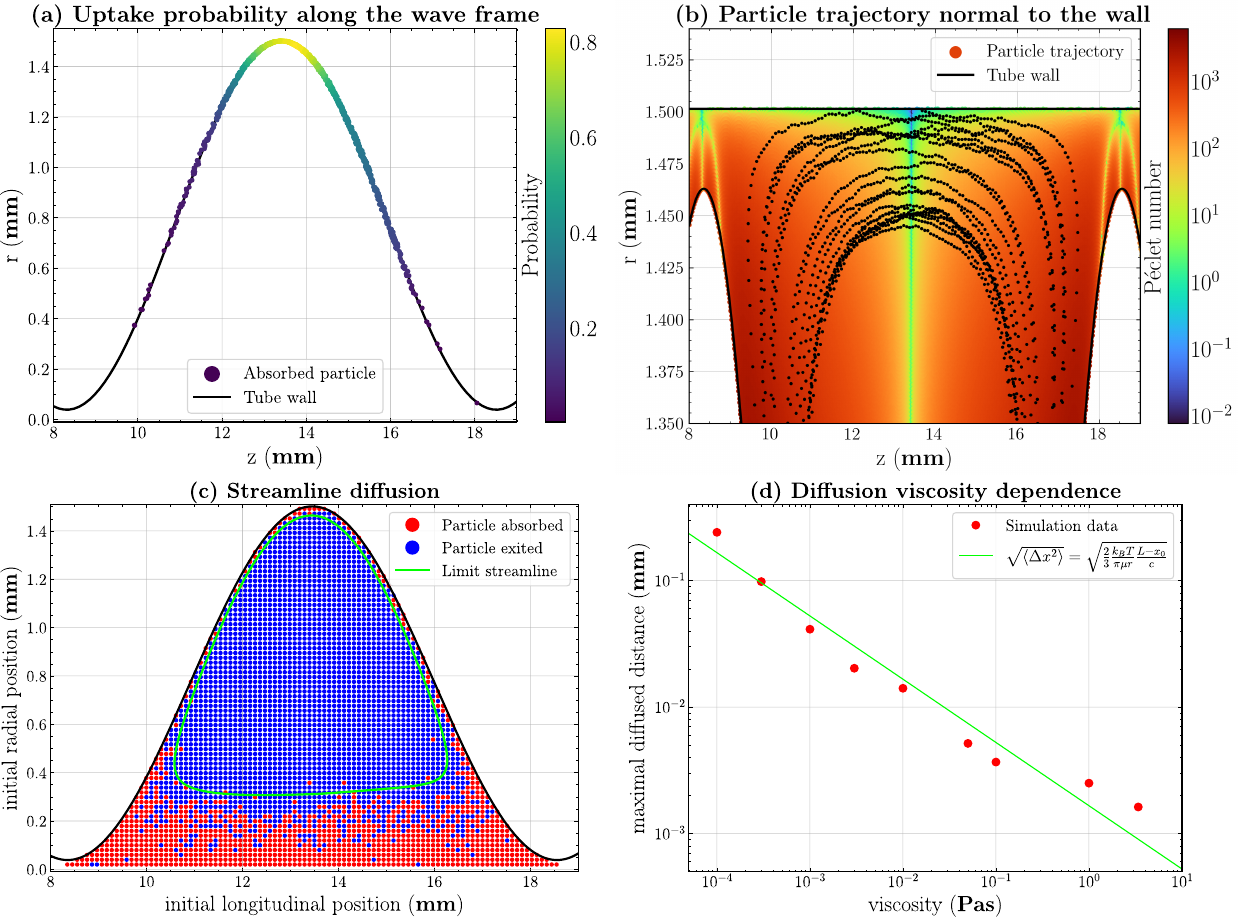}
		\caption{\textbf{Particle uptake in peristalsis is based on increased diffusive behavior around the wave crest resulting in viscosity governed uptake for a finite length tube.}\\
        Uptake probability along the wave frame (a) indicates that particles get mainly taken up at the wave crest. Trajectory of a taken up particle (b) visualized in the wave frame transformed to be normal to the tube wall, reveals that uptake happens in the diffusion dominated, low Péclet number, region close to the crest.
		(c) The limit streamline (green) correctly predicts the threshold of particle being taken up versus exiting at the end of the tube. The limit streamline indicates the largest distance a particle can hop streamlines by diffusion to get taken up before exciting the tube. Simulation data with $\mu = 10^{-3}\SI{}{\pascal\second}$.
        (d) Comparison of predicted and simulated maximal diffused distance reveals a decrease in uptake with fluid viscosity.}
		\label{fig:perist mechanism}
	\end{figure}

		To understand the selective uptake in peristalsis we take a closer look into the underlying mechanism. The uptake probability along the wave frame reveals that particles are mainly taken up near the wave crest, see Fig.~\ref{fig:perist mechanism}a. To understand this localized uptake we map out the trajectory of an exemplary particle, see Fig.~\ref{fig:perist mechanism}b, where for a better visualization of the movement towards/away from the wall, the y-coordinate is transformed such that it is always normal to the wall-shape. Additionally, the Péclet number $Pe = \frac{a_0v}{D}$ is shown in the background to highlight the areas where advection, at large Péclet, or diffusion, at low Péclet, dominates. The trajectory shows that the particle follows the streamline with negligible fluctuations as long as advection dominates over diffusion near the center of the tube indicated by large Péclet numbers, resulting in a straight trajectory. When moving to the crest, however, the particle trajectory starts to fluctuate significantly due to the fluid velocity approaching zero at the crest which leads to a dominance of fluctuations indicated by the low Péclet numbers in the vicinity of the tube crest. Thus, the particle deviates from its initially straight trajectory and jumps between streamlines, which allows it to move towards the wall and be taken up. Moreover, around the wave crest the streamlines are much closer to each other, see Fig.~\ref{fig:perist vs seg}e, which enhances the hopping among streamlines as less distance is required to reach a different streamline. Based on these observations we propose an analytical argument that determines whether a particle is taken up or not as a function of its initial particle position, the time it spends in the tube and the diffusion coefficient which depends on the fluid viscosity. Mechanistically particles can only be taken up if they diffuse across streamlines toward the tube wall within the time they spend in the tube. The maximal travel distance due to diffusion is $\langle\Delta x^2\rangle = 4Dt = 4 \frac{k_BT}{6\pi\mu r}\frac{L-x_0}{c}$, where the travel time $t$ is limited by the time spend in the tube, which itself is determined by the initial longitudinal position $x_0$ relative to the tube length $L$ and the contractions wave speed identical to the bolus speed $c$. Identifying the streamline that is maximally within $\Delta x$ distance to the tube wall or the tube center line everywhere in the wave frame, which we denote the limit streamline, results in a great prediction on the taken up particle's initial position, see Fig.~\ref{fig:perist mechanism}c. This analytical argument makes the effect of viscosity on particle uptake in a peristaltic tube easy to grasp, as $\Delta x$ is inversely proportional to fluid viscosity via its control of particle diffusivity. In order to validate this dependency we perform simulations with different viscosity's, see Fig.~\ref{fig:perist mechanism}d. For each absorbed particle we compute its initial, minimal distance to the wall based on the streamline it is located on. Out of all taken up particles we then plot the largest initial distance which is equivalent to the maximal diffused distance as we are considering only particles taken up within the tube. The results are then compared to the predicted trend in Fig.~\ref{fig:perist mechanism}d. Increasing the fluid viscosity clearly results in a decreasing diffusion distance as predicted by the analytical argument, thus, confirming the inverse proportionality of uptake and fluid viscosity.

    \subsection{Shear thinning improves uptake in segmentation whilst not impeding peristaltic cleaning}
\begin{figure}[h!]
    		\includegraphics[width = \columnwidth]{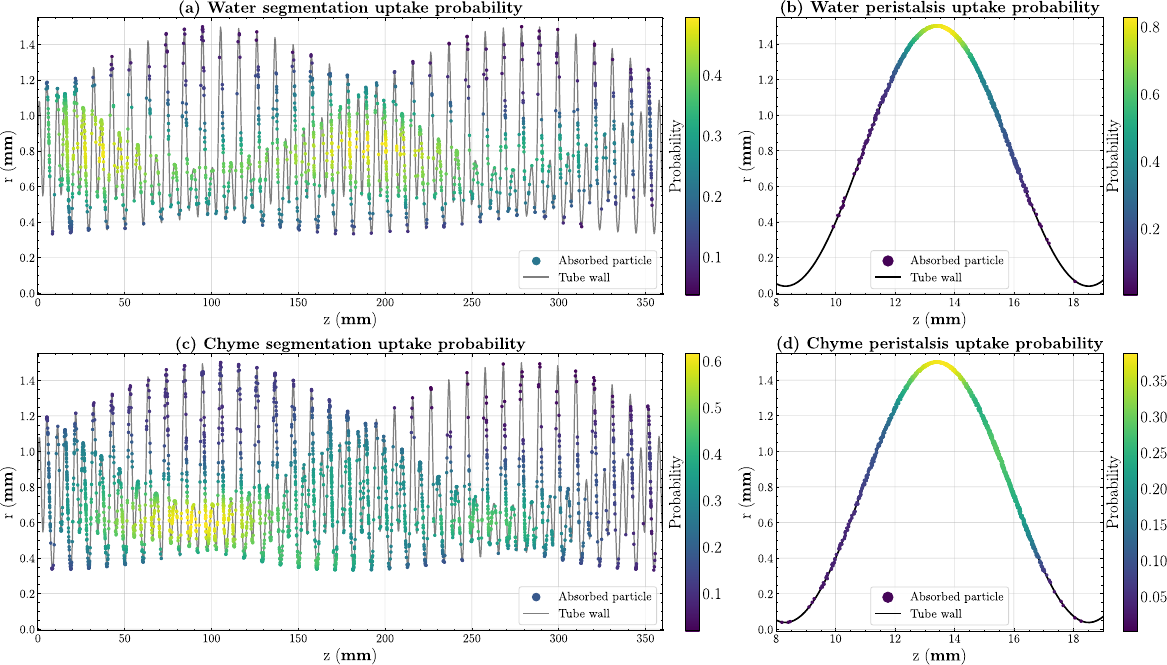}
            \caption{\textbf{A shear-thinning fluid causes a shift in uptake probability towards regions of higher shear rates. This heavily influences uptake in segmentation but only marginally in peristalsis.}\\
            Uptake probability for segmentation with water (a) and shear-thinning chyme (c) retracing uptake positions onto the contraction shape at time $t = \SI{0}{\s}$ reveals that shear-thinning fluid greatly enhances overall particle uptake. 
            Uptake probability for peristalsis with water (b) and shear-thinning chyme (d) mapped onto the wave frame indicates in addition to the peak at the wave crest a second  but marginal peak of uptake for shear-thinning fluid further along the tube wall further toward highest flow velocities at the wave inflection.}
    		\label{fig:seg mechanism}
    	\end{figure}

        Having characterized uptake dynamics in a Newtonian fluid we now turn to a Non-Newtonian fluid. Comparing the uptake location along the wave frame in peristalsis for Newtonian water and Non-Newtonian chyme, see Figs.~\ref{fig:seg mechanism}b and \ref{fig:seg mechanism}d, we observe a second probability peak in addition to the one at the wave crest for the shear thinning fluid to emerge around the inflection point of the wave. Shear-thinning locally decreases the apparent fluid viscosity in zones of high shear rate, which in turn increases the uptake probability as a lower viscosity increases diffusivity. For peristalsis the shear rate is highest at maximal occlusion, however, due to the high occlusion and the resulting shape of the streamlines, see Fig.~\ref{fig:perist vs seg}e, the vast majority of particles does not traverse the region of maximal occulusion, instead they circle towards the wave crest. The few particles which escape the main bolus are either absorbed near the point of maximal occlusion, as can be seen in the bottom left/right of Fig.~\ref{fig:seg mechanism}d, or get trapped in a second bolus with fewer particles, one contraction period downstream the main bolus. Still the shear rate retains large values at the tube wall and only slowly decreases toward the wave crest. Thus, near the wave inflection point the majority of particles are close enough to the wall to experience the increased shear rate causing them to diffuse toward the wall despite a non-zero velocity field. Therefore, in a shear thinning fluid there are two key mechanisms that lead to particle uptake. The first is the diffusive streamline hopping near the wave crest due to low velocities also observed in water. Whereas the second one is the result of the shear-thinning properties as a second uptake area is created due to the high shear-rates along the wall. 
        
        For segmentation it is more difficult to investigate the uptake dynamics as it is not possible to study the uptake probability in the wave frame due to the superposition of waves making such a transformation impossible. However, the contraction pattern still exhibits spatial periodicity, here $\lambda = \SI{357}{\milli\meter}$, which allows for a similar analysis. Mapping for each particle its position, time of uptake and the shape of the surrounding wave onto the same corresponding wave shape of a selected point in time, we extract the uptake probability along the segmentation pattern, see Fig.~\ref{fig:seg mechanism}a and \ref{fig:seg mechanism}c for water and chyme. 
        First considering the uptake behavior of particles in water subjected to segmentation, Fig.~\ref{fig:seg mechanism}a, a clear difference to peristalsis is observed as uptake is not predominantly near the wave crests/points of minimal occlusion. The segmentation contraction pattern is shaped in such a way that there is always a high amplitude contraction followed by a low amplitude contraction. The difference between the two amplitudes then oscillates spatially and at some point, e.g.~at $z\approx\SI{190}{\milli\meter}$ in Fig.~\ref{fig:seg mechanism}a, changes sign such that the low amplitude increases and the high amplitude decreases. Particles are always sloshed back and forth from decreasing to increasing amplitude either moving towards the tube entry or exit depending on their position when the amplitude starts to decrease. Around the point where high and low amplitude are most similar the uptake probability is by far the highest with the general trend being that at low amplitude contractions more particles are taken up than at high amplitude contractions. This makes sense as on the one hand the fluid velocity is much lower in the low amplitude regions, thus, diffusion dominates there compared to the high amplitude waves. On the other hand the region between two adjacent crests is traversed the most by particles when the amplitudes are in-/decreasing the fastest i.e.~at the inflection point where both high and low amplitude are of the same size, thus, increasing the uptake probability there. 

        By changing the fluid properties to be shear-thinning, Fig.~\ref{fig:seg mechanism}c, just like for peristalsis, a shift in the uptake probability is observed towards regions of higher shear rates. For segmentation these are in regions where the amplitude difference reaches its maximum and in regions of maximal occlusion. The probability in the same amplitude region is now significantly lower than for water as the shear rates in segmentation are much higher than in the regions where particles travel during peristalsis causing the shear-thinning to be more influential here. Thus, a key difference between the contraction patterns is observed as the shear dependent uptake causes much higher uptake numbers in segmentation compared to peristalsis due to the contraction shape of the walls. For peristalsis the difference in uptake numbers is only marginal as particles do not cross the high shear regions whereas for segmentation the region of highest shear is exactly the one that is mostly occupied or traversed by particles. Therefore, a shear-thinning fluid maintains the house keeping function of peristalsis, i.e.~particles are mainly flushed out while the the uptake function of segmentation is enhance compared to Newtonian fluids.

\section{Discussion}
    Using finite element simulations a simplified model of the small intestine was investigated. Based on a previous result on intestinal contraction \cite{AgneseMainPaper}, the interplay of contractile driven flow and particle behavior was analyzed. Here particle's diffusion was found to be of key relevance for nutrient uptake in the gut. By contrasting the uptake behavior in a Newtonian fluid with a Non-Newtonian shear thinning fluid, which resembles the chyme found in the intestine \cite{Shearthinningchickendigesta, Shearthinningpigdigesta}, a shift in uptake probability towards regions of high shear rates was observed. This causes the uptake to increase, showing that the secreted, viscous digestive juices e.g.~from the pancreas, are not only important to break down the food into smaller pieces but also to improve the nutrient uptake along the intestine. Here, the mentioned increase in uptake was much greater for segmentation compared to peristalsis where the enhancement was only marginal. Thus, shear thinning properties lead to an increased nutrient uptake whilst at the same time not impeding the function of peristalsis, to remove unnecessary remnants, as here the regions of high shear rates are not traversed by particles due to the bolus formation such that shear-thinning does not enhance uptake.\\
    Comparing the two main contraction patterns, it was confirmed that peristalsis traps particles in a bolus, which itself quickly exits the intestinal track \cite{Jimenezlozano, AgneseMainPaper}. Segmentation, in contrast, causes much longer residence times, promotes spatial distribution and mixing of the fluid particle \cite{AgneseMainPaper, Motilitycomparison}. The latter is especially important given that digestive juices are rather characterized as shear-thinning than Newtonian. In combination with the longer residence times observed for segmentation we confirm segmentation's role as a contraction pattern responsible for postprandial nutrient uptake \cite{Originofsegmentation} in contrast to peristalsis which quickly clears out remnant contents after digestion.\\
    Studying single particle trajectories subjected to peristaltic contractions we discovered an analytical argument to predict particle uptake based on the diffusion across streamlines before exiting the tube. The analytical argument is based on the fact that only near the wave crest the Péclet number is sufficiently low such that diffusion dominates and particles may hop between streamlines such that they come close enough to the tube wall to be taken up. Comparing the numerical results with the analytical prediction a good agreement was found and it was, thus, shown that the amount of particles taken up during peristalsis is inversely correlated to the viscosity of the fluid. This novel finding could be experimentally verified by tracking microsphere particles in a microfluidic devices mimicking physiological gut flows \cite{Ibanez, minigut}\\
    Although our work here considered the physiologically relevant setting of the small intestine, the fluid dynamics insight in how fluid property and contraction pattern control particle uptake versus longitudinal particle transport may well be of relevance across contractile tubes in life and technical applications. 

\section{Acknowledgement}
    We thank Agnese Codutti for stimulating discussions. This work has received funding from the European Research Council (ERC) under the European Union’s Horizon 2020 research and innovation program (grant agreement No. 947630, FlowMem) and the Human Frontier Science Program Organization through Research Grant number RGP0001/2021 to K.~A..

\newpage

\end{document}